\documentclass[fleqn]{article}
\begin{document} 
\title{{\bf Binary Tree Summation Monte Carlo Method for Potts Models}} 
%
%
%
%
%
\author{ 
Jian-Sheng Wang,$^{1,2}$ Oner Kozan,$^2$ and Robert H. Swendsen$^2$ \\
%
%
%
%
{\small
$^1$Singapore-MIT Alliance and Department of Computational Science,}\\
{\small National University of Singapore, Singapore 119260, 
Republic of Singapore.}\\
{\small $^2$Department of Physics, Carnegie Mellon University, 
Pittsburgh, PA 15213.}
} 
\date{}
 
\maketitle              
 
\section{Introduction} 
 
Efforts to develop better and more efficient algorithms for Monte
Carlo simulations have a long history, in which the Fortuin-Kasteleyn
(FK) transformation \cite{Kasteleyn} for Potts models has played a
pivotal role.  The most common use of this transformation has been to
create algorithms in which clusters of spins in the Potts model are
flipped simultaneously \cite{Swendsen-Wang,Wolff}.  In this
paper, we present a new algorithm using the FK representation in the
spirit of recent work by Newman and Ziff \cite{Newman-Ziff} on the
percolation problem.  The method produces independent samples and sums
up a large number of configurations for each sweep.
The partition function and thermodynamic averages for all values of
the temperature $T$ can be computed from a single run.
 
Consider the Potts model with the Hamiltonian, 
\begin{equation} 
  H(\sigma) =  -J \sum_{\langle i,j\rangle} \delta_{\sigma_i, \sigma_j}, 
\end{equation} 
where the summation is over nearest neighbor pair, $\sigma_i = 1, 2,
\ldots, q$.  The FK transformation allows us to write the partition
function in the percolation representation as
\begin{equation} 
  Z = \sum_{\Gamma} p^b (1-p)^{M-b} q^{N_c(\Gamma)},  
\end{equation}  
where we sum over all configurations of bonds connecting nearest
neighbor sites, $p=1-\exp\bigl(-J/(kT)\bigr)$, $b$ is the number of
bonds present, $M = dN$ is the maximum number of possible bonds, and
$N_c(\Gamma)$ is the total number of clusters for a given
configuration of bonds.
 
In order to evaluate thermodynamic averages in this representation, we
carry out the summation over all configurations in two steps. First,
we sum over the number of bonds $b$, and then for each value of $b$,
we sum over all configurations consistent with that number of bonds.
Thus we can write
\begin{equation} 
   Q(p) = Z^{-1} \sum_{b} p^b (1-p)^{M - b} c_b\, Q_b, 
\end{equation} 
where $c_b = \sum_{\Gamma_b} q^{N_c(\Gamma_b)}$ and  
\begin{equation} 
   Q_b = {1 \over c_b} \sum_{\Gamma_b} q^{N_c(\Gamma_b)} Q(\Gamma_b), 
\end{equation} 
where $\Gamma_b$ is a configuration with exactly $b$ bonds.  If we
could compute $c_b$ and $Q_b$ for some observable $Q(\Gamma)$ for
every $b$, then we could compute the function at any values of $p$ or
$T$.
 
The problem then reduces to the computation of the normalization
constants $c_b$ and the expectation values in an ensemble with
probability distribution proportional to $q^{N_c(\Gamma)}$.  For the
special case of $q=1$ (bond percolation), the values of $c_b$ are the
binomial coefficients and the configurations with fixed number of
bonds are weighted uniformly.  For general case, we describe two
sampling methods.  The first is very simple, and useful conceptually,
but exponentially inefficient.  The second turns out to be very
efficient.
 
\section{A Survival and Death Process} 
 
Starting with an empty lattice, one sweep consists of repeated
application of the following steps until the process dies:
 
\begin{enumerate} 
 
\item Pick an unoccupied neighbor pair at random for the next bond. 
 
\item If inserting a bond   
 \begin{enumerate} 
\item does not change the cluster number ($\Delta N_c=0$), accept the   
       configuration; 
\item merge two clusters, so that the cluster number decreases by 1,  
       ($\Delta N_c = -1$), accept the configuration with probability $1/q$, 
        or reject the configuration and terminate the process 
       (and begin the next sweep from an empty lattice). 
  \end{enumerate} 
 
\item Take statistics of the survival configurations (with equal weights). 
\end{enumerate} 
 
The probability distribution of the configurations generated in this
manner with $b$ bonds is
\begin{equation} 
 P(\Gamma_b) = { b! (M-b)! \over M !}  
                q^{N_c(\Gamma_b) - N_c(\Gamma_0)}. 
\end{equation} 
The survival probability after $b$ bonds is then
\begin{equation} 
   S_b = \sum_{\Gamma_b} P(\Gamma_b) = { b! (M-b)! \over M !}  
       \, c_b\, q^{-N}, 
\end{equation} 
where $N$ is the total number of sites.  As the survival probability
decays exponentially with the number of bonds added, we expect very
poor statistics for large values of $b$.  However, the eventual
survival probability, $S_{M}$, is equal to $q^{-N+1}$, which is
nonzero.

\section{Binary Tree Summation Method} 
 
To deal with the exponentially decreasing survival probability, we
have developed the following method for summing multiple bond
sequences in a single sweep.  For a configuration $\Gamma_b$ we count
the number of type-0 bonds $n_0$ of unoccupied pairs of sites 
that are on the same cluster, and the number of type-1 bonds 
$n_1$ of unoccupied pairs of sites that would
connect two different clusters.  Clearly $n_0+n_1+b = M$.  At each
step, we pick a type-1 bond with equal probability from among all
current type-1 bonds.  Starting with an empty lattice and merging
clusters at each step, we continue until all sites are members of the
same cluster.  Along the way, we collect statistics $Q(i)$ for each of
the $N$ configurations generated.
 
After a sweep has been carried out, we construct all possible paths
that a full simulation of both type-0 and type-1 bonds would have
taken if we had joined clusters in exactly the same sequence as in the
actual simulation, but had also inserted a random number of type-0
bonds at each step.  At each choice between types of bond, the type-1
bond is given a relative weight of $n_1/q$ and a type-0 bond $n_0$.
The total weight for a path starting from an empty
lattice to a particular configuration is the product of the factors
$n_0$ or $n_1/q$ depending on the path taken.
 
To understand this algorithm, imagine that we had actually followed
the branching process.  At each step, a configuration may split into
two configurations, one with a type-1 bond added (with probability
$1/n_1$) and one with a type-0 bond added (with probability of
$1/n_0$).  The probability of appearance of a particular path is
$\prod [n_{f(k)}(\Gamma_k)]^{-1}$, where $f(k) = 0$ or 1 depending on
the choice of type-0 or 1 bond.  In taking statistics, we have
weighted with the inverse of the factor, multiplied by additional
factors of $q$ proportional to the number of clusters $N_c(\Gamma)$.  The
net effect is the required sample average with overall weight
$q^{N_c(\Gamma)}$.  Now the key observation is that we do not need an explicit
simulation for the type-0 bonds, since the type-0 bonds have no effect
on the measured quantities.
 
The binary tree summation algorithm has several attractive features.
First, the usual slowing down due to correlation between samples is
absent. Each sweep is independent.  Second, for each sweep, data for a
very large number of samples are collected.  Although they are highly
correlated, an exponentially large number of paths can be summed
efficiently with $O(N^2)$ operations per sweep.  Third, unlike
multicanonical simulations \cite{Berg} or the flat histogram method
\cite{Wang-Swendsen-JSP}, there are no unknown weighting factors to
determine.
 
\section{Implementation}  
Let $w(b,i)$ be the weight of the total contributions from all
possible paths to the state specified by the number of bonds $b$ and 
merge sequence number $i$.  These
quantities can be calculated from the starting condition $w(0,i) =
\delta_{i,0}$, the constraint that $n_0(b,i) \ge 0$, and the recursive
equation
\begin{equation} 
  w(b+1,i) = w(b,i) n_0(b,i) + w(b,i-1) n_1(b,i-1)/q, 
\end{equation} 
where $n_1(b,i) = n_1(i)$, $n_0(b,i) = n_0(i) - b + i$.  The value
$w(b,i)$ is nonzero only for $b \ge i$.  The computation of the
weights $w(b,i)$ is similar to that of binomial coefficients.  The
final contribution to the statistics at $b$ number of bonds is
\begin{equation} 
  Q_b = \langle W_b \rangle^{-1} 
             \Bigl\langle \sum_{i=0}^{N-1} w(b,i) Q(i) \Bigr\rangle,
\end{equation}  
where the average is over simulation sweeps, and $W_b = \sum_i w(b,i)$
is the total weight at a given $b$.  We can then show that
\begin{equation} 
  \langle W_b \rangle = { M! \over (M-b)! } S_b, 
\end{equation} 
which allows us to compute $c_b$.  The result is numerically identical
to computing the conditional survival probability 
$S_{b+1}/S_b$ from the expectation value of $Q = (n_0 + n_1/q)/(M-b)$.
 
During the simulation, the values of $n_0$ and $n_1$ can be updated
efficiently.  For each cluster, we keep a list of unoccupied bonds
with other clusters.  When two clusters (A and B) are joined, we merge
the smaller one with the larger one, and remove and count the number
of bonds $n_{{\rm AB}}$ connecting the clusters.  We update according
to $n_0 \leftarrow n_0 + n_{{\rm AB}} -1$, $n_1 \leftarrow n_1 -
n_{{\rm AB}}$.  The timing of our program shows that this part of the
algorithm scales nearly linearly with number of sites $N$, as
expected.
 
\section{Results and Discussions} 
We note that the coefficients $c_b$ are related to the density of
states $n(E)$, which gives the coefficients of the partition function
polynomial in the variable $\exp\bigl(-J/(kT)\bigr)$.  In the FK
percolation representation, it becomes a polynomial in $p/(1-p)$. By
a proper change of variables, we can find exact results \cite{Beale}
of $c_b$ for the two-dimensional Ising model.
 
\begin{table} 
\caption{Errors for $10^6$ sweeps with respect to the exact results of 
an $L \times L$ Ising model.
While $\epsilon_1$ is taken from an average over many runs, the other
results are from a single run.  The CPU times (on a 1.53GHz Athlon)
are in units of $10^{-6}$ second per sweep per site.}
\renewcommand{\arraystretch}{1.4} \setlength\tabcolsep{10.2pt}
\begin{tabular}{llllll} 
\hline\noalign{\smallskip} 
$L$ & 4 & 8 & 16 & 32 & 50 \\ 
\noalign{\smallskip} 
\hline 
\noalign{\smallskip} 
cpu $t$      & 1.88  &   2.81   &    6.92  &   24.0   &   72.6\\
$\epsilon_1$ & 0.0000634 & 0.000178 &  0.00049 &  0.0032 &  0.031\\
$\epsilon_{{\rm MAX}}^E$ & 0.000128  & 0.000113  & 0.00012  & 0.0015 & 0.0068\\
$\epsilon_{{\rm AVE}}^E$ & 0.0000184 & 0.0000103 & 0.0000055 & 0.000046 & 0.00014\\
$\epsilon_{{\rm MAX}}^C$ & 0.000306  & 0.00046   & 0.00124  & 0.0177 & 0.096\\
$\epsilon_{{\rm AVE}}^C$ & 0.000034  & 0.000031  & 0.000040 & 0.00050 & 0.0020\\
\hline 
\end{tabular} 
\label{Tab1a} 
\end{table} 
 
We define the following errors to test our method against exact
results for the two-dimensional Ising model:
\begin{eqnarray} 
   \epsilon_0 &=&  \bigl|  q^{N-1} c_M/c_0 - 1 \bigr|,  \\ 
   \epsilon_1 &=& {1 \over M} \sum_{b=0}^M \bigl|  
                            {c_b / c_b^{\rm{exact}}} - 1\bigr|,  \\ 
   \epsilon_{{\rm MAX}}^Q &=& \max_{T}\, \bigl| Q(T) - Q^{\rm{exact}}(T)  \bigr|, \\ 
   \epsilon_{{\rm AVE}}^Q &=&  
                  \int_0^1 dx\, \Bigl| Q\bigl(T(x)\bigr) - 
                             Q^{\rm{exact}}\bigl(T(x)\bigr)  \Bigr|, 
 \quad T(x) = {x \over 1-x}. 
\end{eqnarray} 
Due to a special cancellation for this algorithm, $\epsilon_0$ is
exactly zero.  The error $\epsilon_1$, along with the maximum and
average errors in the energy and specific heat, are listed in Table 1.
Since the algorithm asymptotically takes $O(N^2)$ operations per
sweep, a fair comparison with other methods should compare the total
CPU times.  A comparable N-fold way transition matrix Monte Carlo
(TMMC) run took 1.9 microsecond per sweep per site on the same
machine.  Thus, the present method is superior for small lattices with
linear size $L \leq 16$.  For $L=32$ it is comparable to TMMC
\cite{Wang-Swendsen-JSP}.  For much larger lattices, it becomes less
favorable mainly due to the $O(N^2)$ nature of the algorithm.  It
is quite likely that we can speed up the computation using special
properties of the weights.
 
Our method is applicable for Potts models with any number of states,
including fractional or negative values.  Work is currently in
progress to apply this algorithm to a number of problems of interest.

%
 

\begin{thebibliography}{8.} 
\addcontentsline{toc}{section}{References} 
 
\bibitem{Kasteleyn} P. W. Kasteleyn and C. M. Fortuin, J. Phys. Soc.  
Jpn Suppl. {\bf 26}, 11 (1969); C. M. Fortuin and P. W. Kasteleyn,
Physica {\bf 57}, 536 (1972). 
 
\bibitem{Swendsen-Wang} R. H. Swendsen and J.-S. Wang, Phys. Rev. Lett.  
  {\bf 58}, 86 (1987). 
 
\bibitem{Wolff} U. Wolff, Phys. Rev. Lett. {\bf 62}, 361 (1989);
Nucl. Phys. {\bf B322}, 759 (1989).
 
\bibitem{Newman-Ziff} M. E. J. Newman and R. M. Ziff, Phys. Rev. Lett.  
{\bf 85}, 4104 (2000); Phys. Rev. E {\bf 64}, 016706 (2001). 
 
\bibitem{Berg} B. A. Berg and T. Neuhaus,  Phys. Rev. Lett. {\bf 68},  
9 (1992). 
 
\bibitem{Wang-Swendsen-JSP} J.-S. Wang and R. H. Swendsen, J. Stat. Phys. 
{\bf 106}, 245 (2002). 
 
\bibitem{Beale} P. D. Beale, Phys. Rev. Lett. {\bf 76}, 78 (1996).  
 
 
\end{thebibliography}
\end{document}